\documentclass[aps,onecolumn,secnumarabic,nobalancelastpage,amsmath,amssymb,nofootinbib]{revtex4}
\pdfoutput=1

\usepackage{graphicx}% Include figure files
\usepackage{dcolumn}% Align table columns on decimal point
\usepackage{bm}% bold math

\begin{document}
%\setpagewiselinenumbers
%\modulolinenumbers[1]
%\linenumbers

%\preprint{APS/123-QED}

\title{Multiply resonant high quality photonic crystal nanocavities}% Force line breaks with \\

\author{Kelley Rivoire$*$, Sonia Buckley, and Jelena Vu\v{c}kovi\'{c}}

\email{krivoire@stanford.edu} %% email address is required

\address{E. L. Ginzton Laboratory, Stanford University, Stanford, CA 94305-4088}


\begin{abstract}
We propose and experimentally demonstrate a photonic crystal nanocavity with multiple resonances that can be tuned nearly independently. The design is composed of two orthogonal intersecting nanobeam cavities. Experimentally, we measure cavity quality factors of 6,600 and 1000 for resonances separated by 382 nm; we measure a maximum separation between resonances of 506 nm. These structures are promising for enhancing efficiency in nonlinear optical processes such as sum/difference frequency and stimulated Raman scattering.
\end{abstract}

\maketitle
%\clearpage
Photonic band gap nanocavities confine light into subwavelength volumes with high quality factor, and have been used to build ultracompact optoelectronic devices such as lasers\cite{notomi_laser} and modulators\cite{notomi_modulator}, as well as to study fundamental physics such as cavity quantum electrodynamics\cite{dirk_reflectivity,imamoglu}. State of the art photonic crystal nanocavity designs\cite{noda_nmat, notomi_1D, parag_highQ, noda_billionq} can be optimized to generate quality factors exceeding one million for a single cavity resonance. For nonlinear optical interactions such as frequency conversion or stimulated Raman scattering, however, it is desirable to have multiply resonant nanostructures\cite{johnson} with arbitrary frequency separation and with good spatial field overlap. While many nanocavity designs have multiple resonant modes within a single photonic band gap, it is difficult to independently control their frequencies; moreover, field patterns of different resonances typically have minimal spatial overlap, and the absolute frequency separation between resonances is limited by the size of the photonic bandgap. Photonic crystals and quasicrystals\cite{bouwmeester} can have multiple photonic band gaps, but the size of the higher order band gaps greatly diminishes for finite thickness structures; in addition, in such planar structures, higher order band gaps are located above the light line, implying that the resonances at those frequencies would also have low Q factors, resulting from the lack of total-internal reflection confinement. Band gaps for different polarizations (e.g. transverse magnetic and transverse electric) have also been proposed\cite{burgess, yinan_tetm, mccutcheon_broadband} to generate additional resonant modes; however, it is difficult to independently tune the frequencies, and the proposed designs require relatively thick membranes which are more difficult to fabricate. Here, we propose and experimentally demonstrate a crossed nanobeam photonic crystal cavity design that allows at least two individually tunable resonances with a frequency separation larger than the size of the photonic bandgap in a single nanobeam.

The proposed multiply resonant structure is shown in Fig. 1a, 1b. The basic element of the design is the nanobeam\cite{notomi_1D, ippen, mccutcheon_sin, lalanne, sauvan, parag_highQ}, a 1D periodic photonic crystal waveguide clad in the other two directions by air. Our design uses two orthogonal intersecting beams to achieve lithographically tunable resonances at multiple frequencies.
A cavity is formed in each beam by introducing a central region with no holes (cavity length $l$) and tapering the lattice constant and hole size near the cavity region\cite{lalanne, parag_highQ}. In each beam, confinement along the periodic direction is provided by distributed Bragg reflection; confinement out of plane is provided by total internal reflection. Confinement in the in-plane direction orthogonal to the beam axis is provided by total internal reflection, and in the case of beams with overlapping photonic band gaps, also by distributed Bragg reflection (as in structures designed for minimizing crosstalk in waveguide intersections \cite{Johnson:98}). This structure allows nearly independent tuning of each resonant frequency by tuning the parameters (e.g. width, lattice constant, cavity length) of each beam. Additionally, the structure has natural channels for coupling through each beam to an access waveguide at each wavelength\cite{qimin}. To optimize the design, parameters of cavity length, lattice constant ($a$), number of taper periods $N$, distance between holes in the taper region ($a_i$, $i=1,2,...N$), hole size in the taper region ($d_i$, $i=1,2,...,N$),  and beam width ($w$) were varied in each beam. Fig. 1c shows the 3D finite difference time domain (FDTD)-simulated $E_y$ field pattern for a resonance at 1.55 $\mu$m, with mode volume 0.35$(\lambda/n)^3$ where $n$ is the refractive index, and Q=19000, limited by loss in the vertical direction. Fig. 1d shows the field pattern of $E_x$ for a resonance primarily localized by the vertical beam with wavelength 1.1 $\mu$m (also limited by vertical loss) and mode volume $0.47(\lambda/n)^3$. The shorter wavelength mode likely has a lower quality factor in part due to its narrower width than the horizontal beam and partly due to FDTD discretization error (12 points per period). Decreasing the vertical lattice constant further (10 points per lattice period) leads to resonances with separation of 574 nm (1550 nm and 976 nm).

Because a real structure has a fixed membrane thickness, decreasing the periodicity in one beam relative to the other leads to larger relative thickness $t/a$, redshifting the wavelengths of the bands; therefore achieving resonances with relative frequency $f_2/f_1$ requires superlinear scaling of the feature size, e.g. $a_2/a_1 > f_2/f_1$, as shown in Fig. 2a (there is a second less important contribution because refractive index increases for higher frequencies, from 3.37 to 3.53 for range plotted in Fig. 2a). Fig. 2b shows tuning of the quality factor of an individual resonance calculated by FDTD by adjusting the cavity length (a minimum cavity length is required to avoid overlap between the holes forming cavities in each beam). The resonant wavelength as well as quality factor of the beam can also be tuned by varying other parameters; Fig. 2c shows the change in resonant frequency and quality factor as width of the horizontal beam is varied. To study the overall effect f varying a single parameter, we plot a figure of merit for nonlinear frequency conversion $FOM=Q_1Q_2/\sqrt{V_1V_2}$ in Fig. 2d (right axis), which is seen to depend primarily on $Q_2$ (as mentioned previously, for large horizontal beamwidths, the quality factor of the higher frequency mode is limited by diffraction into the orthogonal beam).

For nonlinear frequency conversion applications, it is important for cavity field patterns to have large spatial overlap. Defining the nonlinear overlap, normalized to 1, as\cite{yinan_tetm}
\begin{equation}
\gamma \equiv \frac{\epsilon_{NL} \int_{NL}dV \sum_{i,j,i\neq j} E_{1,i} E_{2,j} } {\sqrt{\int dV \epsilon |E_1|^2} \sqrt{\int dV \epsilon |E_2|^2}}
\end{equation}
where $NL$ indicates nonlinear material only, we calculate $\gamma$=0.02 for the structures shown in Fig. 1. This number could be increased to 0.07 by decreasing the number of taper periods from 5 to 3 to further localize the field to the central region; however, the quality factors are reduced to 1440 and 1077 respectively, although this could likely be increased by reoptimization of other parameters.

A scanning electron microscope (SEM) image of a fabricated structure is shown in Fig. 3a. The structures are defined by e-beam lithography and dry etching, as well as wet etching of a sacrificial AlGaAs layer underneath the 164 nm thick GaAs membrane. To compare the experimental structure with the proposed design, we simulate the fabricated structure by converting the SEM image to the binary refractive indices of air and GaAs\cite{dirk_sem}. The thresholded SEM image used for simulation is shown in Fig. 3b. Figs. 3c, 3d show the simulated electric field for the two resonances of the fabricated structure, with expected resonances at 1477 nm (Q=4100) and 1043 nm (Q=540). Finally, we experimentally characterize our design by performing a reflectivity measurement in the cross-polarized configuration\cite{rivoire_gap} (Figs. 3e, 3f) using light from a tungsten halogen lamp which is linearly polarized using a Glan Thompson polarizer and polarizing beamsplitter. The cavity is oriented at 45 degrees to both the polarization of the incident light and orthogonal polarization used for measurement. We measure quality factors of 6600 at 1482.7 nm and 1000 at 1101 nm. Differences between simulation of fabricated structures and experimentally measured resonances are most likely due to computational error in binary thresholding of the refractive index. Experimentally, the largest separation between resonances we have measured is 506 nm.

In conclusion, we have proposed and demonstrated the design of a photonic crystal cavity with multiple nearly independent resonances, with up to 382 nm separation measured experimentally and Q$>$1000 for both resonances. The design also features independent waveguide channels for two resonances, which facilitates their spatial separation, as is desirable in many applications. Our design is promising for resonantly enhancing on-chip nonlinear frequency interactions, such as stimulated Raman scattering (which could be achieved in semiconductors such as GaAs or Si by a slight detuning between the parameters of the two beams) and sum/difference frequency generation. By choosing the parameters of the two beams to be the same, these structures could also serve as polarization-degenerate cavities, which would be desirable for coupling to spin states of embedded quantum emitters\cite{skolnick} or for building polarizaton entangled photon sources based on a bi-excitonic cascade from a single quantum dot\cite{benson_yamamoto, calzone}.

Financial support was provided by the National Science Foundation (NSF Grant ECCS-10 25811 ).  KR and SB supported by Stanford Graduate Fellowships and the NSF GRFP (SB). This work was performed in part at the Stanford Nanofabrication Facility of NNIN supported by the National Science Foundation under Grant No. ECS-9731293.

\begin{figure}[h]
\includegraphics[width=8.5cm]{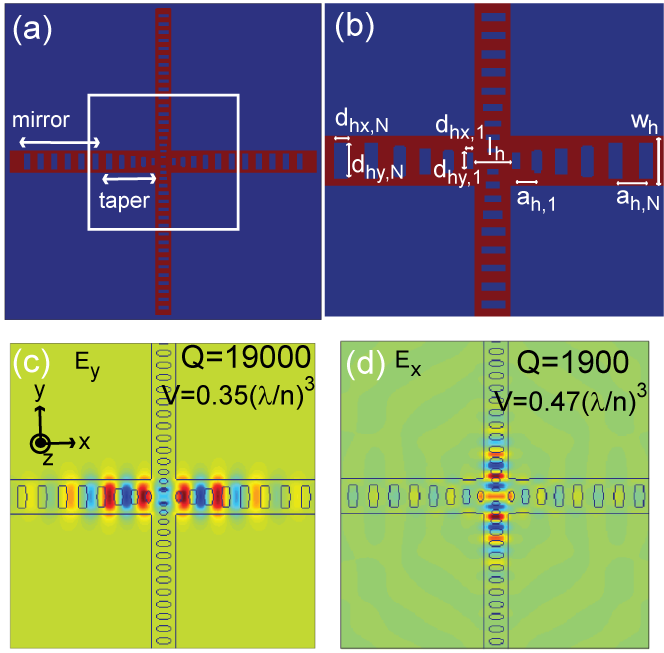}% Here is how to import EPS art
\caption{\label{fig:figure1}
  (a) Illustration of cavity design, showing intersecting orthogonal nanobeams with taper and mirror regions, as well as central cavity. (b) Detail of white box in (a). Parameters used to form resonance, shown for cavity in horizontal (subscript h) beam: $l_h$ indicates cavity length; $d_{hx,N}$ and $d_{hy,N}$ indicate hole sizes in mirror region; $d_{hx,1}$ and $d_{hy,1}$ indicate hole sizes in first taper period; $a_{h,N}$ indicates periodicity in mirror region; $a_{h,1}$ indicates periodicity in first taper period, $w_h$ indicates beam width. The corresponding parameters are similarly introduced for the vertical beam (with subscript $v$). The thickness of both beams (in the z direction) is $t$. Parameters are changed linearly inside the taper. (c) Field pattern of $E_y$ for cavity mode localized by horizontal beam. Parameters are: $a_{h,N}$=453 nm, $a_{v,N}$=272 nm, $d_{hx,1}/d_{hx,N}=d_{hy,1}/d_{hy,N}$=0.5, $a_{h,1}/a_{h,N}=a_{v,1}/a_{v,N}$=0.7, $l_h/a_{h,N}=1.2$, $l_v/a_{v,N}$=0.83, $w_h/a_{h,N}$=1.65, $w_v/a_{v,N}$=1.8, $d_{hy,N}/w_h=d_{vx,N}/w_v$=0.7, $d_{hx,N}/a_h=d_{vy,N}/a_{v,N}$=0.5, refractive index $n=3.37$, with slab thickness $t/a_{h,N}$=0.35, $N$=5, and 6 mirror periods for both beams. Resonant wavelength is 1.55 $\mu$m with Q=19,000 and V=0.35$(\lambda/n)^3$. (d) Field pattern of $E_x$ for cavity localized by vertical beam. $n=3.46$ and other parameters same as in (c). Resonant wavelength is 1103 nm with Q=1900 and V=0.47$(\lambda/n)^3$. }
\end{figure}

\begin{figure}[h]
\includegraphics{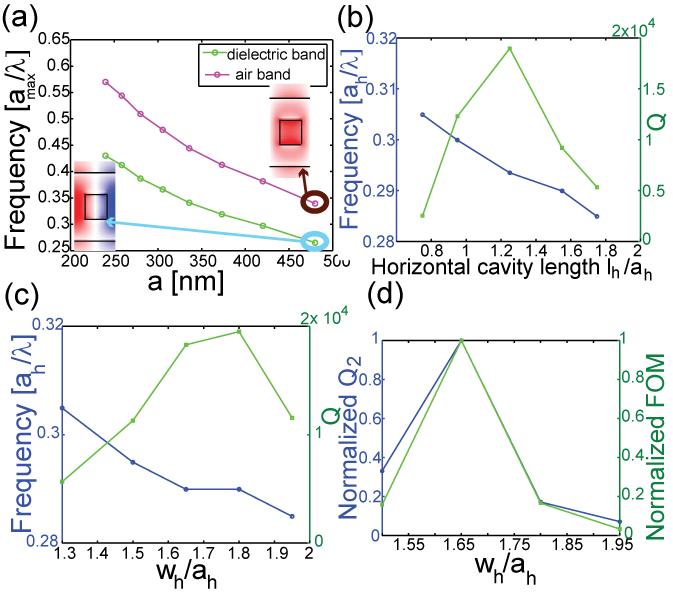}% Here is how to import EPS art
\caption{\label{fig:figure3}
 (a) Frequencies of dielectric and air bands of a single GaAs nanobeam as lattice constant is varied. Simulations are performed using plane wave expansion with supercell approach (MIT MPB)\cite{mpb}. Parameters are $w/a$=1.65, $h_y/w$=0.6, $h_x/a$=0.5, t=160 nm, $a_{max}$=450 nm. $n$ varies from 3.37 to 3.53 in the simulated frequency range. (b) Change in cavity resonant wavelength and Q in crossed beam structure for cavity mode localized in horizontal beam as cavity length is varied. Maximum Q occurs for $l_h/a_h$=1.25. (c) Change in horizontal (long wavelength) cavity resonant wavelength and Q in crossed beam structure as $w_h$ is varied. (d) Change in vertical beam quality factor and frequency conversion figure of merit $FOM=Q_1Q_2/\sqrt{V_1V_2}$ as a function of $w_h$.}
\end{figure}

\begin{figure}[h]
\includegraphics{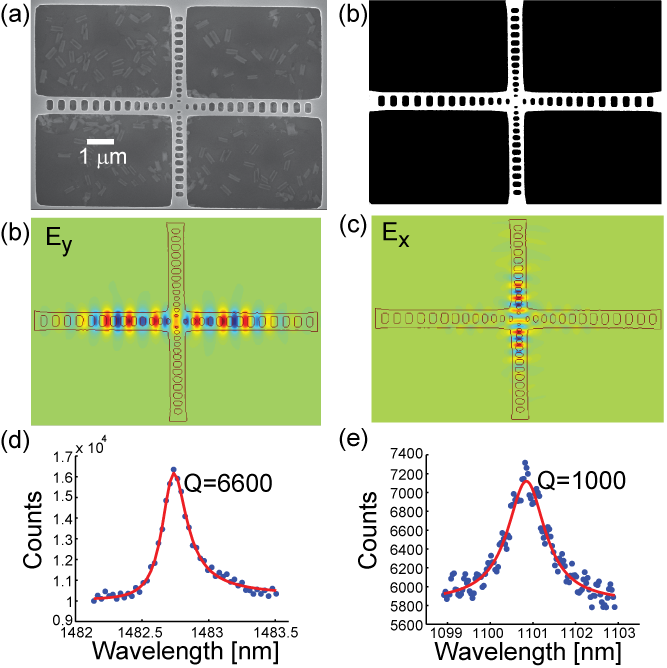}% Here is how to import EPS art
\caption{\label{fig:figure3}
  (a) SEM image of cross-beam structures fabricated in 164 nm thick GaAs membrane with $a_{h,N}$=470 nm and $a_{v,N}$=320 nm fabricated in 164 nm membrane. (b) Thresholded binary image of SEM used for simulating fabricated structure. (c) FDTD simulation of $E_y$ for cavity resonance at 1477 nm. (d) Simulated $E_x$ for cavity resonance at 1043 nm. (e) Reflectivity measurement of cavity mode at 1482.7 nm with Q=6600. (f) Reflectivity measurement of cavity resonance at 1101 nm with Q=1000.}
\end{figure}

\end{document}